\documentclass[aps,pra,groupedaddress,twocolumn,showpacs]{revtex4-1}
\usepackage{graphics}
\usepackage{graphicx}
\usepackage{longtable}

\begin{document}

\title{Anisotropic $LMN$ dielectronic resonances from ratios of
magnetic-dipole lines}

\author{Yu. Ralchenko}
\email[Electronic mail: ]{yuri.ralchenko@nist.gov}
\author{J.D. Gillaspy}
\affiliation{
National Institute of Standards and Technology, Gaithersburg, Maryland
20899-8422
}

\date{\today}
\begin{abstract}

Signatures of multi-keV $LMN$ dielectronic resonances in highly-charged $3d^n$
ions of tungsten were detected in the intensity ratios of extreme-ultraviolet
magnetic-dipole lines within ground configurations. The measurements were
performed with an electron beam ion trap at beam energies of about 6 keV.
Large-scale collisional-radiative modeling incorporating magnetic sublevels of
autoionizing levels showed the significance of anisotropy effects due to the
monodirectional propagation of the electron beam. The observation method allows
simultaneous resolved registration of dielectronic resonances from several
ions.

\end{abstract}

\pacs{32.30.-r,32.80.Zb,34.50.Fa}
\maketitle

\section{Introduction}

Dielectronic recombination (DR) \cite{Burgess,GriemBook,Kunze} is one of the
important processes affecting ionization balance and related parameters (e.g.,
radiative power losses) of plasmas involving highly-charged ions (HCI). DR is
a two-step process involving dielectronic capture (DC), that is, excitation of
the atomic electron with simultaneous capture of the incident electron, and
radiative stabilization of autoionizing states produced during DC. Although a
particular resonant state can only be produced with an electron of specific
energy, a Maxwellian plasma has electrons of all energies and thus the DR rate
is never zero for thermal plasmas. Due to a weak dependence of the DC rates on
the ion charge $z$ and strong $\sim z^4$ dependence of the radiative rates,
radiative stabilization is more probable for highly-ionized atoms. That is why
DR is often the most dominant recombination mechanism for HCI plasmas,
especially at low densities. In addition to its effect on the ionization
distribution, DR frequently results in the appearance of strong satellite
lines that may serve as reliable diagnostics of plasma properties. 

Dielectronic resonance states have been extensively studied with storage rings
\cite{Schi11,Muller09,Orban09}, electron beam ion traps (EBITs)
\cite{Knapp89,Knapp91,DeWitt92,Ba3d,Naka06,Beilmann2010}, tokamaks
\cite{Urnov07,Bitter03}, and other devices \cite{Laser12}. The DR measurements
with EBITs are primarily based on two techniques. One method utilizes
extraction of ions at different beam energies and analysis of their abundances
\cite{Ali91,DeWitt92,Naka06}. It is not unusual, however, that the measured
ionization distributions differ from the originally produced abundances inside
the trap, and thus additional corrections may be required \cite{Zhang2010}.
Another method is based on EBIT beam ramping \cite{Knapp89}. In this case, a
relatively high beam energy, higher than the ionization potential of a
specific ion, is used to make the ion under study as abundant as possible.
This energy is typically much higher than the energy required to produce
outer-shell dielectronic resonances. Then the beam energy is quickly changed
to a lower value (or series of values) and data is collected before the
ionization balance has time to change substantially.  The beam energy is then
restored to its initial value and held there for a relatively long time in
order to restore the initial charge balance.  This process is repeated over
many duty cycles in order to integrate the signal acquired during the brief
parts of the cycle that the beam is at the low energies. 

Elementary energy balance considerations show that the dielectronic resonances
due to excitation of the outer-shell electrons cannot be produced with the
beam energy larger than the ionization energy of the ion. It is possible,
however, to create resonances due to excitation of inner-shell electrons. For
instance, in Ref.~\cite{Ba3d} the LMM resonances (i.e., inner L electron 
excited into the M shell with the free electron captured into the M shell)
were identified and analyzed in the x-ray spectra from the Ba ions with an
open $3d$ shell, Ba$^{34+}$ and Ba$^{35+}$. Note that the ionization
potentials of Ba$^{34+}$ and Ba$^{35+}$ are 2142~eV and 2256~eV, respectively
\cite{ASD}, while the beam energies were about 2400~eV. Good agreement was
found between theory and experiment for the ratios of DR to radiative
recombination cross sections. 

In this paper we report in situ measurements and analysis of the $LMN$
dielectronic resonances in about 50-times ionized atoms of tungsten with an
open $3d$ shell. Tungsten is currently a leading candidate for the
plasma-facing material in the divertor region of ITER \cite{Hawryluk09}, and
therefore its spectroscopic properties are being actively studied in many
laboratories \cite{KramidaW}. While theoretical papers on DR of tungsten are
numerous (see, e.g., \cite{SafW1,LiW,SafW2,ShenW,BeharW} and references
therein), the experimental data are rather scarce. Over the last decade,
dielectronic resonances in tungsten were measured in I-like W$^{19+}$
\cite{Schi11} and in Si-like to N-like ions \cite{Bie09}. The latter paper
reported only the unresolved overlapping $LMn$ ($n$ = 3 to 10) resonances
measured in the x-ray range.

Under typical steady-state EBIT conditions, the balance between the neighbor
ion stages is determined by the corresponding ionization and recombination
processes:

\begin{equation}
\frac{N_z}{N_{z+1}} = \frac{\tilde{R}_{rr}+\tilde{R}_{cx}+\tilde{R}_{dr}}{\tilde{R}_{ion}},
\label{eq1}
\end{equation}
where $N_z$ and $N_{z+1}$ are the total populations of the ions with charges
$z$ and $z$+1, and $\tilde{R}$'s are the total rates for radiative
recombination (rr), charge exchange (cx) with the neutrals in the trap,
dielectronic recombination (dr), and ionization (ion). Equation (\ref{eq1})
assumes that multi-electron processes are negligible. For highly-charged
high-Z ions the double charge exchange (CX) may actually not be much weaker
than the single-electron process; however, the lack of data on double
CX in approximately 50-times ionized atoms prevents one from taking it into
account. 

Since DR is a resonant process, $\tilde{R}_{dr}$ is typically zero for a
nearly monoenergetic EBIT beam. However, when the beam energy $E_b$ overlaps
with the resonant energies, $\tilde{R}_{dr} \neq 0$ and the ionization balance
may change drastically. Here we propose to measure multi-keV inner-shell
dielectronic resonances from low-energy spectroscopic footprints of the
changing ionization balance in the EBIT. To demonstrate this, we use the
intensity ratios of forbidden magnetic-dipole (M1) lines from the ground
configurations of neighbor ions. Several dozen M1 lines from all $3d^n$ ions
of tungsten were recently identified in the extreme ultraviolet (EUV) range
from 10~nm to 20~nm \cite{M1}. It was also found that many of those
well-resolved lines have a rich diagnostic potential for applications in
fusion plasmas. Moreover, we found that the low-lying M1 levels within the
ground configurations are primarily populated by the radiative cascades
following excitation from the lowest ground level. Since in low-density
plasmas (including EBITs) almost all ion population is in its ground state,
the M1 line intensities are in effect proportional to the total ion
populations. Accordingly, their ratio is expected to be sensitive to
modifications of the ionization balance. This new method to detect
dielectronic resonances with the M1 line ratios is described below.

\section{Experiment} 

The EBIT at the National Institute of Standards and Technology (NIST)
\cite{Gillaspy_1997} was used to produce and trap highly charged ions of
tungsten. The low charge ions were injected into the trap using a metal vapor
vacuum arc (MEVVA) \cite{MEVVA}, and then ionized to the required degree. The
beam energies in the present experiment varied between approximately 5500 eV
and 6400 eV, and the beam current was 100 mA. 

The EUV spectra from tungsten ions were recorded with a flat-field
grazing-incidence spectrometer that has been described elsewhere
\cite{Blagojevic_2005}. A gold-coated concave grating with variable spaced
grooves was used in combination with a liquid-nitrogen-cooled back-illuminated
charge-coupled device (CCD) array having a matrix of 2048 $\times$ 512 pixels.
The tungsten emission was recorded between 10~nm and 20~nm with a resolving
power of approximately 350. As in our previous measurements \cite{M1}, this
resolution was sufficient to identify most of the observed lines. Each
spectrum consisted of 10 one-minute exposures, and spurious signals from
cosmic rays were removed. 

\begin{figure}
\includegraphics[width=0.45\textwidth]{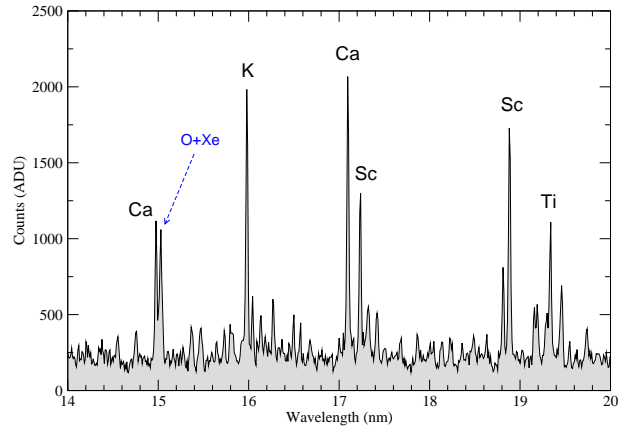}
\caption{\label{fig1} 
Measured spectrum (in analog-to-digital units of CCD) at the nominal beam
energy of 6057 eV. The strongest lines of tungsten are identified by their
isoelectronic sequences. A strong impurity line at 15.02~nm due to emission
from oxygen and xenon ions is shown by an arrow.}
\end{figure}

One of the measured spectra between 14~nm and 20~nm at the nominal beam energy
$E_b$ = 6057 eV is shown in Fig.~\ref{fig1}. The strongest lines of the W
ions are the 15.962-nm line in K-like $3d$ ion, the 14.959-nm and 17.080-nm
lines in Ca-like $3d^2$ ion, the 17.216-nm and 18.867-nm lines in Sc-like
$3d^3$, and the 19.319-nm and 19.445-nm lines in Ti-like $3d^4$ ion \cite{M1}.
Each line in the measured spectra was fit with a Gaussian function and a
constant background, and the derived intensities (areas) were used to obtain
the line intensity ratios. Since it is the line intensities rather than
wavelengths that are of interest in the present work, even an approximate
calibration of the measured spectra that provides a one-to-one correspondence
with the known wavelengths is sufficient here.

\section{Modeling} 

As mentioned above, the $LMn$ dielectronic resonances are produced when the
$L$ electron ($2p_{3/2}$, $2p_{1/2}$, or $2s_{1/2}$ in relativistic notation)
is excited into the $M$ shell and a free electron is captured into an atomic
shell with the principal quantum number $n$. The value of $n$ under EBIT
conditions can be estimated in the limit of a very high ion charge $z$ when
atomic structure becomes quasi-hydrogenic. In units of ($z$+1)$^2$Ry (Ry
$\approx$ 13.6057 eV is the Rydberg constant), the energy difference between
$n$=2 and $n$=3 is 5/36, and the ionization potential for the $n$=3 shell is
IP$_3$=1/9. Numerous EBIT studies show that the beam energy required to make a
specific ion abundant is on the order or higher than its ionization potential.
Therefore, if a free electron has the energy of approximately IP$_3$, then
during dielectronic capture for $L$-$M$ excitation it plunges 1/36(=5/36-1/9)
below the ionization potential, that is, into $n$=6. Certainly, the energy
structure for a 50+-times ionized atom of tungsten cannot be well described by
the exact hydrogenic formulas and the electron energy is normally larger than
the ionization potential but the above estimate of $n\sim6$ is quite close to
reality, as will be seen below.

One of the widely used parameters representing the contribution of a
particular dielectronic resonance to DR is the dielectronic resonance strength
defined here as

\begin{equation}
R_\alpha = \frac{g_\alpha}{g_i}
\frac{A^a_{\alpha;i}\sum_{f}{A^{rad}_{\alpha;f}}}{\sum_{j}{A^a_{\alpha;j}}+\sum_{f}{A^{rad}_{\alpha;f}}},
\label{eq2}
\end{equation}
where $g_\alpha$ is the statistical weight of the doubly excited state
$\alpha$ in the recombined ion $z$, $g_i$ is the statistical weight of the
initial state $i$ in the recombining ion $z+1$, $A^{a}_{\alpha;j}$ is the
autoionization rate from $\alpha$ to state $j$, and $A^{rad}_{\alpha;f}$ is
the radiative rate from $\alpha$ to state $f$. The summation indices $f$ and
$j$ enumerate states in the recombined and recombining ions, respectively.
Note that the summation index $f$ in Eq. (\ref{eq2}) extends over both
autoionizing and non-autoionizing states. It is customary to exclude the
latter from the sum in the numerator, however, for highly-charged high-Z ions
the radiative rates become stronger than the autoionization rates. Therefore
radiative transitions between autoionizing states would  likely result in
decays into non-autoionizing states and provide additional contribution to the
total dielectronic recombination. Therefore we retain the autoionizing states
in the $f$-summation in the numerator.  Then, in the general case, more than
one state in the recombining ion may contribute to dielectronic capture into a
state $\alpha$. However, due to the relatively strong M1 transitions in
highly-charged high-Z ions the populations of the lowest excited states of
$3d^n$ are much smaller than that of the ground level, and therefore we
neglect their contribution to $R_\alpha$ in Eq.(\ref{eq2}). Finally, it should
be mentioned that the resonance strength only {\it approximately} shows how
much a specific doubly excited state contributes to dielectronic
recombination; the accurate answer can only be obtained with a detailed
collisional-radiative (CR) modeling \cite{PFR} which is provided below. 

\begin{figure}
\includegraphics[width=0.45\textwidth]{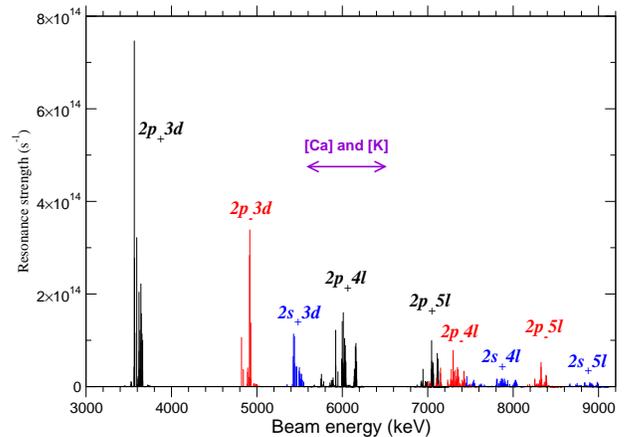}
\caption{\label{fig2} 
Calculated resonance strengths (in s$^{-1}$) for the $LMn$ ($n$=3--5)
resonances in Ca-like W. Arrow shows the approximate range of beam energies where both
Ca-like and K-like ions are abundant. The labels show the quantum numbers for
the L-electron and the recombined free electron. For instance, $2s_{+}3d$
corresponds to the $2s2p^63s^23p^63d^3$ configuration.}
\end{figure}

Large-scale CR modeling requires the generation of extensive sets of atomic
data, such as energy levels, radiative and autoionization rates, and
electron-impact cross sections. The data used in the present work were
calculated with the relativistic-model-potential Flexible Atomic Code (FAC)
\cite{FAC}. This code is widely used for highly-charged high-Z
ions, and its accuracy has proven to be sufficient for such atomic systems.
The calculated $LMn$ ($n$=3--5) resonance strengths (Eq.~(\ref{eq2}))for
dielectronic capture from the K-like ion of tungsten into the Ca-like ion are
presented in Fig.~\ref{fig2} as a function of $E_b$. The resonances
corresponding to different excitations from the $n$=2 shell are given in
different colors: black for $2p_{3/2}$ (often denoted as $2p_+$), red for
$2p_{1/2}$ ($2p_-$), and blue for $2s_{1/2}$ ($2s_+$). Figure~\ref{fig2}
shows that the energy difference between the subshells of $n$=2 is
non-negligible, so that $\Delta E(2p_{3/2}-2p_{1/2}) \approx$ 1~keV and
$\Delta E(2p_{3/2}-2s_{1/2}) \approx$ 2~keV. This deviation from the
quasi-hydrogenic approximation should certainly modify the  $n \approx$ 6
estimate. Then, in addition to the well-known decrease of resonance strengths
with $n$, one can notice that the lowest resonances do not overlap and thus
their contribution to DR can be pinpointed with the narrow spread of energies
in the EBIT beam. Finally, and most importantly, the range of the beam
energies, where both K-like and Ca-like ions are observed to be most abundant
\cite{M1} (horizontal arrow in Fig.~\ref{fig2}), overlaps with the
$L_{3/2}MN$ resonances at approximately 6~keV. Hence, one can expect that the
measurements of ionization balance between the K-like and Ca-like ions near
6~keV should indicate presence of such dielectronic resonances.

The CR modeling used here differs from the approach implemented in our
previous papers on $3d^n$ ions \cite{M1,M1Hf} where only singly- and
doubly-excited states below the first ionization potential were included.
Since L$_{3/2}$MN dielectronic recombination is the primary subject of this
study, all autoionizing states $2s^22p^5_{3/2}3s^23p^63d^3$ and
$2s^22p^5_{3/2}3s^23p^63d^24l$ are added to the model. In the absence of
two-electron processes, the ionization balance is determined only by
ionization and recombination between two ion states, and therefore the CR
model included only Ca-like and K-like ions of tungsten. Even within a two-ion
model the total number of fine-structure levels was about 10,500 since in
addition to approximately 1300 autoionizing states in the Ca-like ion one has
to account for all possible autoionization channels. 

The electron-impact collisional data were calculated with FAC and fit using
simple formulas with the correct asymptotic behavior, and a database was
created for simulations with the non-Maxwellian code NOMAD~\cite{NOMAD}.
Unlike the previous studies on M1 lines \cite{M1,M1Hf}, we do not group the
atomic levels with high principal quantum number $n\geq$4 into ``superterms";
this is done to provide the most accurate representation of all processes
affecting the observed line emission and populations of autoionizing states. 

Generally, CR simulations for EBITs may utilize a number of free parameters
that can be fixed using the experimental data. Thus, the unknown CX rate was
deduced from the spectra measured at non-resonant beam energies
($\approx$6400~eV) and then used in calculations for all other energies. This
fit procedure partially takes into account the higher-order CX processes. The
beam electron energy distribution function (EEDF) was taken to be a Gaussian
with full width at half-maximum of 50~eV which sets the energy resolution. It
is not uncommon in EBIT modeling to use a rectangular EEDF for spectra
simulations due to the smooth energy dependence of collisional cross sections.
However, for the present analysis a Gaussian beam provides a better agreement
with the observed shapes of spectral features. Finally, the effective beam
energy that is seen by ions in the trap is known to be smaller than the
nominal value due to space charge effects. Accordingly, the beam energies in
the following discussion and figures are reduced by 100~eV from their nominal
values to match the calculated positions of resonances.

\section{Comparisons and discussion}

\begin{figure}
\includegraphics[width=0.45\textwidth]{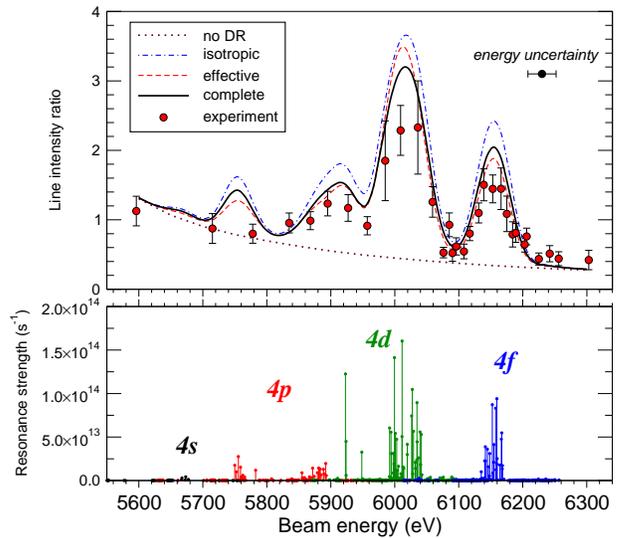}
\caption{ 
Top: circles -- measured intensity ratio of the 17.080~nm line in W$^{54+}$
and the 15.962~nm line in W$^{55+}$; dot-dashed line -- simulation with
$m$-averaged autoionizing levels (AL) assuming isotropic collisions; dashed
line -- simulation with $m$-averaged AL and effective anisotorpic corrections;
solid line --  simulation with $m$-resolved AL; dotted line -- simulation
without AL/DR. Both the energy uncertainty of $\pm$22~eV for all data points
and the line ratio uncertainties correspond to $2\sigma$. Bottom: resonance
strengths for 786 autoionizing states $2s^22p^53s^23p^63d^24l$ in W$^{54+}$.}
\label{fig3} 
\end{figure}

The measured intensity ratio of the 17.080-nm line in Ca-like W$^{54+}$ and
the 15.962-nm line in K-like W$^{55+}$ is shown in Fig.~\ref{fig3}(top).
The vertical error bars correspond to the expanded uncertainty at the level of
confidence of 95 \% and are derived from the uncertainties of Gaussian fits of
the measured line profiles using the coverage factor of $k$=2
\cite{NISTGuide,Guide}. Simultaneous with our measurements in W, a small
amount of argon was continuously introduced into the EBIT through the EUV
spectrometer and used to monitor changes in the absolute value of the electron
beam energy through the position of the x-ray radiative recombination (RR)
line into the ground state of H-like argon. These measurements were performed
with a wide-band high-purity Ge solid-state detector. With the exception of a
few energies (which may have involved line blends) the position of the RR line
varied linearly with the nominal electron beam energy, with a standard
deviation of 11~eV, consistent with statistical uncertainties in the fits to
the RR line centers. The error bars in the horizontal direction (beam energy)
in the plots of the present manuscript are taken to be 22~eV (i.e., $k$=2).
Using the data at a nominal beam energy of 5.815~keV, we used the absolute
value of the RR line center and the known value of the ionization energy of
He-like Argon to infer (119$\pm$14)~eV (at the level of 1$\sigma$) for the
space charge correction to the beam energy, in good agreement with the value
of 100~eV obtained from the EUV line ratios.

Without dielectronic resonances, the experimental Ca/K line ratio should be
monotonically decreasing with $E_b$. This dependence is confirmed by the CR
calculation that excluded autoionizing states (dotted line in
Fig.~\ref{fig3}(top)). The measured line ratio, however, clearly shows
presence of resonant features, with the strongest ones near 6020~eV and
6160~eV. The origin of these peaks is clarified in the bottom panel of
Fig.~\ref{fig3} which is  the enlarged part of Fig.~\ref{fig2}
representing different $4l$ channels. The strongest measured peaks are seen to
correspond to the highest-$l$ channels, $4d$ and $4f$. The contribution from
$4p$ near 5880 eV is visibly smaller, and the weakest $4s$ resonances were
missed on the experimental grid.

The non-Maxwellian CR modeling with autoionizing states for the Ca/K line
ratios was performed at several levels of complexity. The dot-dashed line in
Fig.~\ref{fig3}(top) presents the 17.080/15.962 line ratio calculated with
the model that included resonant states as J-resolved atomic levels. Here the
DC rates were obtained from the corresponding autoionization rates by
integration of DC cross sections over the beam EEDF. This approach is
equivalent to the assumption of isotropic collisions without any account of
the monodirectional nature of the EBIT beam. Whereas the dot-dashed curve in
Fig.~\ref{fig3}(top) reproduces the relative intensities of the resonances
reasonably well, it overestimates their magnitudes. This discrepancy may
result from the anisotropic nature of beam propagation and the ensuing effect
on the DC rates. Due to the conservation of the angular momentum projection,
not all magnetic sublevels of the final recombined state can be populated via
dielectronic capture with monodirectional electrons. Accordingly, the net
DC/DR rates are expected to become smaller and thus reduce the disagreement
between modeling and measurements.

The polarization effects in spectral line emission due to anisotropic
excitation and/or dielectronic capture by monodirectional beam electrons are a
subject of continuing interest (see, e.g.,
\cite{PhysRevA.54.1342,PhysRevA.74.022713,Hakel,GillApJ,ORourke,Matula}). To
include the anisotropic effects, one can develop an extensive CR model
resolved in magnetic sublevels for autoionizing states; this model will be
described below. Another approach to this problem can be based on the
introduction of effective corrections that account for non-statistical
populations of the magnetic sublevels. Such corrections can be developed in
several ways; here we use the following multiplicative modification factors
for all autoionization rates and DC cross sections:

\begin{equation}
\xi_{\alpha} =
\frac{A_\alpha^{rad}+A_\alpha^a}{DC_{\alpha;1}}
\sum_{m=-j}^{j}\frac{DC_{\alpha,m;1}}
{A_\alpha^{rad}+A_{\alpha,m}^a},
\label{eqxi1}
\end{equation}
where $A_\alpha^a$ is the total autoionization rate for the state $\alpha$
with the total angular momentum $j$, $A_\alpha^{rad}$ is its total radiative
rate, $DC_{\alpha,1}$ is the isotropic DC cross section from the next ground
state into $\alpha$ (calculated from $A_\alpha^a$ using the detailed balance
principle), $DC_{\alpha,m;1}^a$ is the DC cross section from the next ground
state into the sublevel with the magnetic quantum number $m$, and $
A_{\alpha,m}^a$ is the total autonization rate from the same sublevel.  This
correction can be justified as follows. The states that contribute most to DR
are those with large radiative transition probabilities $A_\alpha^{rad} \gg
A_\alpha^a$. In this case, the population kinetics of the autoionizing states
becomes similar to the coronal model where a level is populated only by
excitations (dielectronic capture in our case) and depopulated only via
radiative decay. The spectral line intensity for coronal condition does not
depend on the radiative rate and is determined by the excitation rate only.
Reformulating this for the present case, the contribution of a particular {\it
strong} dielectronic resonance to dielectronic recombination does not depend
on the radiative stabilization rate but rather on the dielectronic capture
rate only. For such resonances, the correction $\xi_{\alpha}$ becomes simply
the ratio $\sum_m{DC_{\alpha,m;1}}/DC_{\alpha}$. This multiplicative factor
applied to both autoionization rates and DC rates effectively modifies only
the latter ($A_a$ is too small to compete with $A_{rad}$) and is expected to
bring the total DR rate closer to the magnetic-sublevel-resolved model
results. Note also that correction (\ref{eqxi1}) implies that the population
of the recombining ground state is much larger than the populations of other
excited states which, as mentioned above, is a good aproximation
for the present experiment.

The FAC code allows calculation of DC cross sections into and autoionization
rates from the $m$ states, and such a calculation was performed here. The
result of the ``effective" CR calculation with the $\xi_\alpha$ factors is
presented by the dashed curve in Fig.~\ref{fig3}(top). The agreement with
the measured line ratios is seen to improve, especially for the $4f$
resonances and the $4p$ resonances near 5900~eV. 

The most extensive CR model implemented in the present work splits all 1314
autoionizing levels $2s^22p^5_{3/2}3s^23p^63d^3$ and
$2s^22p^5_{3/2}3s^23p^63d^24l$ into magnetic sublevels, bringing the total
number of such states to 9360, and the total number of all states for a
two-ion CR model to more than 18,000. All possible radiative, autoionization
and dielectronic capture rates connecting the $m$-states were calculated with
FAC and used in NOMAD simulations. The 17.080/15.962 line ratio calculated
with this complete model is presented by the thick solid line in
Fig.~\ref{fig3}(top). The ``complete" and ``effective" models show a
comparable level of agreement for $4p$ and $4f$ resonances, but for the
strongest $4d$ resonances near 6020~eV the complete model is closer to the
measured data. These simulations give a strong indication that anisotropic
effects in EBITs are important for analysis of dielectronic recombination. In
particular, this shows that the DR/DC cross sections measured in EBITs cannot
be treated as the isotropic cross sections.

A natural question arises whether monodirectional propagation of the beam
electrons should affect not only dielectronic resonances but also the M1 lines
and their ratios. The important difference between the autoionizing states and
the low excited $3d^n$ levels is that the former are populated due to direct
interaction of beam electrons with an ion while the latter are populated due
to radiative cascades from the excited higher states. While excitations from
the ground state are generally polarization-sensitive, the radiative cascades
smear out polarization signatures for the lowest levels. This was explicitly
shown in the recent study of electric-quadrupole and magnetic-octupole lines
in Ni-like W$^{46+}$~\cite{PhysRevA.81.012505}. 

\begin{figure}
\includegraphics[width=0.45\textwidth]{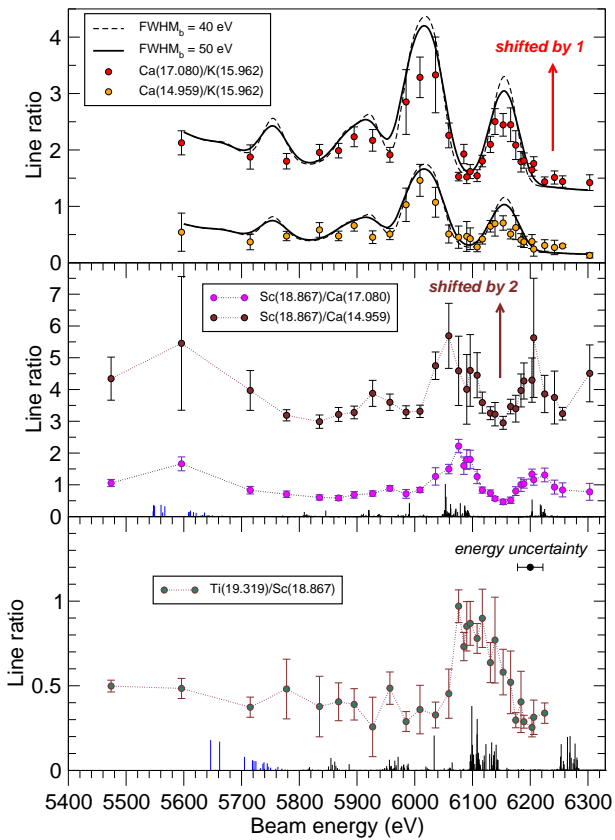}
\caption{\label{fig4} 
Measured line ratios for the Ca-like to K-like ions (top), Sc-like to Ca-like
ions (middle), and Ti-like to Sc-like  ions (bottom). The theoretical ratios
calculated with the anisotropic collisional-radiative model are shown by solid
(FWHM = 50~eV) and dashed (FWHM = 40~eV) lines in the top panel. The
calculated relative strengths for the L($2s_{1/2}$)MM (blue) and 
L($2p_{3/2}$)MN (black) resonances are shown by vertical lines in the middle
and bottom panels. The dotted lines are shown for an eye guide. Both the
energy uncertainty of $\pm$22~eV for all data points and the line ratio
uncertainties correspond to $2\sigma$.}
\end{figure}

A significant advantage of the experimental method presented here consists in
its ability to provide simultaneous {\it resolved} measurements of 
dielectronic resonances from several ions. Even for the modest resolution
spectrometer used here, the magnetic-dipole lines from different ions are
normally well separated in an EUV spectrum, as exemplified in
Fig.~\ref{fig1} (see also Ref.~\cite{M1,M1Hf}), and therefore line ratios
for several ion pairs can be reliably determined from a single EBIT run. Such
analysis was performed here as well, and Fig.~\ref{fig4} shows the
measured ratios for the three pairs of ions, i.e., Ca-like to K-like (top),
Sc-like to Ca-like (middle), and Ti-like to Sc-like (bottom). The
17.080/15.962 ratio (same as in Fig.~\ref{fig3}) and 18.867/17.080 ratio
in the top and middle panels, respectively, are shifted upward by one and two
units, rspectively, for a better visibility. While two other lines, at 17.216
nm in the Sc-like ion and at 19.445 nm in the Ti-like ion, look strong in
Fig.~\ref{fig1}, we did not use them in this analysis as both are blended
by impurity lines that affect their intensities.

Figure~\ref{fig4}(top) shows the experimental 14.959/15.962 Ca/K ratio  
and the ``complete" anisotropic simulation with the beam FWHM of 50 eV (solid
line). This calculation does not include any free parameters that have not
already been fixed in simulation of the 17.080/15.962 ratio. For the
14.959/15.962 ratio, the agreement between theory and measurements is quite
good over the whole range of energies although for both ratios our simulations
slightly overestimate the DR contribution for the $4f$ resonances at 6150~eV.
In addition to the 50-eV-FWHM calculation, we present the 40-eV-FWHM results
(dashed curve) as well. The difference between the two theoretical simulations
is not very significant, however smaller or larger values of the EBIT beam
width were found to be in worse agreement with the measurements. Obviously,
detailed measurements of DR resonances, especially isolated ones such as the
resonance at $\sim$5925~eV in Fig.~\ref{fig3}, can be used to determine
the beam width more accurately.

The available computational facilities did not allow us to perform a
full-scale collisional-radiative modeling for the Sc/Ca or Ti/Sc line ratios,
similar to the results presented for the Ca/K ratios. The total number of
levels in a CR model rapidly increases with the number of electrons $n$ in the
ground configuration $3d^n$. Therefore, in the middle and bottom panels of
Fig.~\ref{fig4} we only present the calculated relative dielectronic
resonance strengths which are shown by the blue ($2s_{1/2}$ inner electron
excitation, $LMM$ resonances) and black ($2p_{3/2}$ inner electron excitation,
$LMN$ resonances) vertical lines. This comparison shows that for the $3d^3$
and $3d^4$ ions of tungsten both positions and strengths of the resonances are
calculated as accurately as for the $3d^2$ Ca-like ion (Figs.~\ref{fig2}
and \ref{fig3}). 

The measured line ratios indicate that the beam energy required to produce a
particular $LMN$ resonance decreases with the increase of the ion charge $z$.
As is seen in Fig.~\ref{fig4}, the energy of the strongest peak due to
$2p_{3/2}4d$ resonances is 6020~eV for Ca-like W$^{54+}$, about 6070~eV for 
Sc-like W$^{53+}$, and 6100~eV for Ti-like W$^{52+}$. Similar dependences on
$z$ are also visible for the weaker $4f$ peak. Such behavior would seem to
contradict the well-known increase of atomic energies with $z$. The
explanation can be derived from a simple analysis of energy balance. The
resonant beam energy is equal to the difference between the excitation energy
$\Delta E_{23}$ from n=$2$ to n=$3$ and the ionization energy IP$_4$ of the
$4l$ subshell. Since excitation involves a deeply screened $2l$ electron,
$\Delta E_{23}$ varies with $z$ much less than IP$_4$ which approximately
increases as $z^2$. These dependences are confirmed by our FAC calculations
along isoelectronic sequences. As a result, the resonant beam energies for the
$LMN$ resonances in $3d^n$ ions of tungsten weakly decrease with the ion
charge.

\begin{figure}[t]
\includegraphics[width=0.50\textwidth]{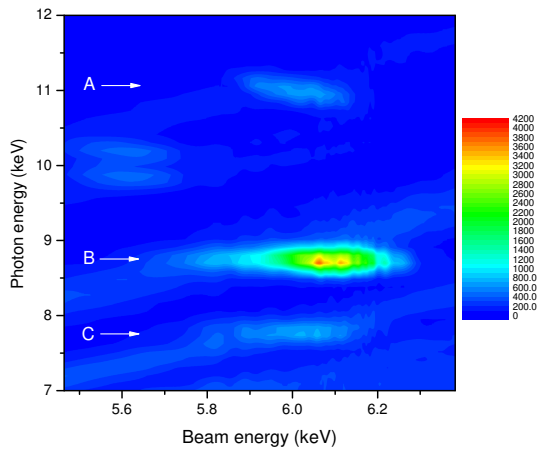}
\caption{\label{fig5} Measured x-ray emission (in arbitrary units)
 between 7.5 keV and 12 keV registered with a Ge solid state 
 detector. A: $2p$-$4l$
 transitions, B: $2p$-$3d$ transitions, C: $2p$-$3s$ transitions.}
\end{figure}

Among the possible radiative stabilizing channels for the $LMN$ autoionizing
states, the strongest ones are due to electric-dipole transitions into the
$2p_{3/2}$ hole, that is, $2p_{3/2}$-$4l$ (referred to below as A) with
$l$=$s$ or $d$, $2p_{3/2}$-$3d$ (B), and $2p_{3/2}$-$3s$ (C). These
transitions result in emission of photons with the energies of about 11~keV,
8.8~keV, and 7.8~keV, respectively. To record these x-rays, we used the
high-purity Ge solid state detector. The measured spectrum presented in
Fig.~\ref{fig5} unambiguously reveals the three radiative channels A, B, and
C for the electron beam energies between 5.7~keV and 6.3~keV. Note that there
is a background of radiative recombination with some RR channels visible as
weak diagonal bands. Unlike the EUV line ratios that are resolved for
individual ions, the stabilizing x-ray photons may simultaneously originate
from several ions that are abundant at a specific beam energy. For instance,
as can be seen from Fig.~\ref{fig4}, dielectronic resonances can be
produced for both Ca-like and Ti-like ions at the beam energy of 6140~eV.
Since our present CR model is restricted only to two ions, it is not currently
possible to accurately analyze the observed x-ray emission. Yet some
characteristics of the measured spectrum confirm our identifications. For
instance, the B transitions extend to the beam energies higher than 6.1~keV
while there are no A transitions in that region. This agrees with the
preferential population of the $4f$ subshell at highest energies as $4f$
electrons cannot decay into $2p$ via electric-dipole transitions.

We mentioned above that the energies of the $2p$-$3l$ inner-shell transitions
(channels B and C) are almost independent of the ion charge. Indeed, the
measured central energies for the B and C features in Fig.~\ref{fig5} do not
visibly change with the beam energy although the ion populations vary
significantly. The changing central energy of the $2p$-$4l$ spectral group A
is, however, clearly discernible as confirmed by the Gaussian fits of this
feature. One can see that the lowest-energy $2p$-$4d$ x-ray photons (from
lower-$z$ ions) appear at higher beam energy and vice versa. This agrees with
the above discussed $z$-dependence of the resonant beam energy and the
movement of the $2p$-$4d$ peak (see Fig.~\ref{fig4}). 

\section{Conclusions}

In this work we proposed a new in situ method of measuring the $LMN$
dielectronic resonances in $3d^n$ ions of heavy elements through intensity
ratios of magnetic-dipole lines within ground configurations. Such resonances
were experimentally observed for Ca-, Sc-, and Ti-like ions of tungsten in an
electron beam ion trap with beam energies of about 6~keV. In addition to the
EUV spectra, x-ray photons due to the radiative stabilization phase of
dielectronic recombination were recorded using a high-purity Ge detector. A
detailed non-Maxwellian collisional-radiative model was developed to simulate
the observed line intensity ratios. It was shown that the isotropic CR
simulations overestimate the effect of DR on the measured line ratios. We
proposed effective corrections to account for non-statistical populations of
magnetic sublevels of autoionizing states and also developed a complete
magnetic-sublevel-resolved CR model. Both new models were found to improve
agreement with the measured line ratios thereby pointing out the importance of
anisotropic effects for dielectronic recombination in EBITs. The remaining
difference between the simulations and experiment may stem from imperfection
of atomic data, incompleteness of the CR model, or for instance multi-electron
processes. These issues should be addressed in future works. It is rather safe
to claim, however, that the measured line ratios offer benchmark data for both
atomic structure theories and advanced non-Maxwellian models of plasma
emission.

\begin{acknowledgments}

Supported in part by the Office of Fusion Energy Sciences of the U.S.
Department of Energy. We are grateful to J.J. Curry for assitance during the
experimental run and to D.~Osin, A.~Kramida, Y.~Podpaly, J. Reader, and
J.N.~Tan for valuable discussions.

\end{acknowledgments}

\bibliography{paper}

\end{document}